\documentclass[aps,prc,amsfonts,superscriptaddress,twocolumn,showpacs,nofootinbib]{revtex4-1}  
\usepackage{graphicx}  
\usepackage{amssymb}   
\usepackage{amsmath}
\usepackage{hyperref}
\usepackage{verbatim}
\usepackage{multirow}

\begin{document}
\title{High-precision $Q_{EC}$-value measurement of the superallowed $\beta^+$ emitter $^{22}$Mg and an {\it ab-initio} evaluation of the $A=22$ isobaric triplet}
\author{M.P.~Reiter}\affiliation{TRIUMF, 4004 Wesbrook Mall, Vancouver, British Columbia, V6T 2A3, Canada}\affiliation{II. Physikalisches Institut, Justus Liebig Universit\"at Giessen, 35392 Giessen, Germany}
\author{K.G.~Leach}\email{kleach@mines.edu}\affiliation{Department of Physics, Colorado School of Mines, Golden, Colorado, 80401, USA}
\author{O.~M.~Drozdowski}\affiliation{TRIUMF, 4004 Wesbrook Mall, Vancouver, British Columbia, V6T 2A3, Canada}\affiliation{Institute for Theoretical Physics, Universit\"at Heidelberg, Philosophenweg 12, D-69120 Germany}
\author{S.R.~Stroberg}\affiliation{TRIUMF, 4004 Wesbrook Mall, Vancouver, British Columbia, V6T 2A3, Canada}
\author{J.D.~Holt}\affiliation{TRIUMF, 4004 Wesbrook Mall, Vancouver, British Columbia, V6T 2A3, Canada}
\author{C.~Andreoiu}\affiliation{Department of Chemistry, Simon Fraser University, Burnaby, British Columbia, V5A 1S6, Canada}
\author{C.~Babcock}\affiliation{TRIUMF, 4004 Wesbrook Mall, Vancouver, British Columbia, V6T 2A3, Canada}
\author{B.~Barquest}\affiliation{TRIUMF, 4004 Wesbrook Mall, Vancouver, British Columbia, V6T 2A3, Canada}
\author{M.~Brodeur}\affiliation{Department of Physics, University of Notre Dame, Notre Dame, IN 46556 USA}
\author{A.~Finlay}\affiliation{Department of Physics and Astronomy, University of British Columbia, Vancouver, British Columbia, V6T 1Z1, Canada}\affiliation{TRIUMF, 4004 Wesbrook Mall, Vancouver, British Columbia, V6T 2A3, Canada}
\author{M.~Foster}\affiliation{Department of Physics, University of Surrey, Guildford, GU2 7XH, United Kingdom}\affiliation{TRIUMF, 4004 Wesbrook Mall, Vancouver, British Columbia, V6T 2A3, Canada}
\author{A.T.~Gallant}\affiliation{Department of Physics and Astronomy, University of British Columbia, Vancouver, British Columbia, V6T 1Z1, Canada}\affiliation{TRIUMF, 4004 Wesbrook Mall, Vancouver, British Columbia, V6T 2A3, Canada}\affiliation{Physical Life Sciences Directorate, Lawrence Livermore National Laboratory, Livermore, CA 94550, USA}
\author{G.~Gwinner}\affiliation{Department of Physics and Astronomy, University of Manitoba, Winnipeg, Manitoba, R3T 2N2 Canada}
\author{R.~Klawitter}\affiliation{TRIUMF, 4004 Wesbrook Mall, Vancouver, British Columbia, V6T 2A3, Canada}\affiliation{Max-Planck-Institut f\"ur Kernphysik, Saupfercheckweg, D-69117 Heidelberg, Germany}
\author{B.~Kootte}\affiliation{Department of Physics and Astronomy, University of British Columbia, Vancouver, British Columbia, V6T 1Z1, Canada}\affiliation{TRIUMF, 4004 Wesbrook Mall, Vancouver, British Columbia, V6T 2A3, Canada}
\author{A.A~Kwiatkowski}\affiliation{TRIUMF, 4004 Wesbrook Mall, Vancouver, British Columbia, V6T 2A3, Canada}\affiliation{Cyclotron Institute, Texas A\&M University, College Station, Texas, 77843, USA}
\author{Y.~Lan}\affiliation{Department of Physics and Astronomy, University of British Columbia, Vancouver, British Columbia, V6T 1Z1, Canada}\affiliation{TRIUMF, 4004 Wesbrook Mall, Vancouver, British Columbia, V6T 2A3, Canada}
\author{D.~Lascar}\affiliation{TRIUMF, 4004 Wesbrook Mall, Vancouver, British Columbia, V6T 2A3, Canada}
\author{E.~Leistenschneider}\affiliation{Department of Physics and Astronomy, University of British Columbia, Vancouver, British Columbia, V6T 1Z1, Canada}\affiliation{TRIUMF, 4004 Wesbrook Mall, Vancouver, British Columbia, V6T 2A3, Canada}
\author{A.~Lennarz}\affiliation{TRIUMF, 4004 Wesbrook Mall, Vancouver, British Columbia, V6T 2A3, Canada}\affiliation{Institut f\"ur Kernphysik, Westfalische Wilhelms-Universit\"at, D-48149 M\"unster, Germany}
\author{S.~Paul}\affiliation{TRIUMF, 4004 Wesbrook Mall, Vancouver, British Columbia, V6T 2A3, Canada}
\author{R.~Steinbr\"ugge}\affiliation{TRIUMF, 4004 Wesbrook Mall, Vancouver, British Columbia, V6T 2A3, Canada}\affiliation{Cyclotron Institute, Texas A\&M University, College Station, Texas, 77843, USA}
\author{R.I.~Thompson}\affiliation{Department of Physics and Astronomy, University of Calgary, Calgary, Alberta, T2N 1N4, Canada}
\author{M.~Wieser}\affiliation{Department of Physics and Astronomy, University of Calgary, Calgary, Alberta, T2N 1N4, Canada}
\author{J.~Dilling}\affiliation{TRIUMF, 4004 Wesbrook Mall, Vancouver, British Columbia, V6T 2A3, Canada}\affiliation{Department of Physics and Astronomy, University of British Columbia, Vancouver, British Columbia, V6T 1Z1, Canada}
\date{\today}

\begin{abstract}

A direct $Q_{EC}$-value measurement of the superallowed $\beta^+$ emitter $^{22}$Mg was performed using TRIUMF's Ion Trap for Atomic and Nuclear science (TITAN).  The direct ground-state to ground-state atomic mass difference between $^{22}$Mg and $^{22}$Na was determined to be $Q_{EC}=4781.40(22)$~keV, representing the most precise single measurement of this quantity to date.  In a continued push towards calculating superallowed isospin-symmetry-breaking (ISB) corrections from first principles, {\it ab-initio} shell-model calculations of the $A=22$ IMME are also presented for the first time using the valence-space in-medium similarity renormalization group formalism. With particular starting two- and three-nucleon forces, this approach demonstrates good agreement with the experimental data.
\end{abstract}

\maketitle
High-precision measurements of nuclear decay properties have proven to be a critical tool in the quest to understand possible physics beyond the Standard Model (BSM)~\cite{Nav13}.  Superallowed $0^+\rightarrow0^+$ nuclear $\beta$-decay data are among the most important to these tests, as they currently provide the most precise determination of the vector coupling strength in the weak interaction, $G_V$~\cite{Har15,PDG16}.  This is only possible in this unique electroweak decay mode, since the transition operator that connects the initial and final $0^+$ states is independent of any axial-vector contribution to the weak interaction.  In fact, the up-down element of the Cabibbo-Kobayashi-Maskawa (CKM) quark-mixing matrix, $V_{\mathrm{ud}}$, is the most precisely known ($0.021\%$)~\cite{PDG16}, and relies nearly entirely on the high-precision superallowed $\beta$-decay $ft$ values determined through measurements of the half-life, decay $Q$-value, and branching fraction of the superallowed decay mode~\cite{Har15}.
\newline\indent
In order to obtain the level of precision required for Standard Model tests, corrections to the experimental $ft$-values must also be made to obtain nucleus-independent ${\cal F}t$ values,
\begin{eqnarray}
{\cal F}t\equiv ft(1+\delta_R)(1-\delta_C)=\frac{2\pi^3\hbar^7\ell n(2)}{2G_V^2m_e^5c^4(1+\Delta_R)},
\label{Ft_value}
\end{eqnarray}
where $\delta_R$ is a transition-dependent radiative correction, $\Delta_R$ is a transition-independent radiative correction, and $\delta_C$ is a nucleus-dependent isospin-symmetry-breaking (ISB) correction.  Although relatively small ($\sim1\%$), these corrections are crucial due to the very precise ($\leq0.1\%$) experimental $ft$ values~\cite{Har15}.  The uncertainty on $G_V$, and consequently $V_{ud}$, is presently dominated by the precision of these theoretical corrections, specifically $\Delta_R$ and $\delta_C$.  With a value of 2.361(38)\%~\cite{Mar06}, the largest fractional uncertainty of any individual correction term is due to the transition-independent radiative correction, $\Delta_R$.  Despite the large uncertainty, the QED formalism that is used in the calculation of this quantity is well understood, suggesting that the central value is accurate.  This situation is not as clear for the ISB corrections however, which have a similarly large uncertainty contribution in the extraction of $G_V$, but require complex nuclear-structure calculations on a case-by-case basis~\cite{Sat09,Tow10b}.

The current extraction of $G_V$ and $V_{ud}$ from the superallowed data uses the shell-model-calculated ISB corrections of Towner and Hardy (TH).  This is largely due to the impressive efforts towards experimental testing~\cite{Bha08,Mel11,Par15} and guidance~\cite{Lea13a,Lea13b,Mol15} that their formalism has been exposed to.  However, as the experimental $ft$ values have become increasingly more precise - particularly in the last decade - the model-space truncations~\cite{Tow10} and small deficiencies that exist in the TH formalism~\cite{Mil08} need to be investigated further.  Perhaps the most important future work may result from efforts towards quantifying any overall model-dependent uncertainty or possible shifts in the $\delta_C$ central values, which still remain elusive due to the extreme complexity of this phenomenological approach to the nuclear shell model.

With increasing computational power, more exact theoretical treatments which were out of reach during the early superallowed reviews, have been under investigation for the past 10 years~\cite{Sat09,Lia09,Aue09,Rod12,Sat16}.  So far, these new methods have provided useful insight into where some of the older phenomenological approaches may be incomplete~\cite{Tow10}, but have not yet reached the level of refinement needed for testing the Standard Model.  These new approaches are nonetheless intriguing, as they may offer some insight into quantifying any elusive model-dependent uncertainties, particularly using {\it ab-initio} many-body approaches based on nuclear forces from chiral effective field theory ($\chi$-EFT)~\cite{Epel09RMP,Mach11PR,Car16}.  These efforts are critical due to the dramatic implications of a deviation from unity in the top-row sum of the CKM matrix resulting from a shift in the $\delta_C$ calculations~\cite{Sat09}.
\begin{figure}[t!]
\includegraphics[width=\linewidth]{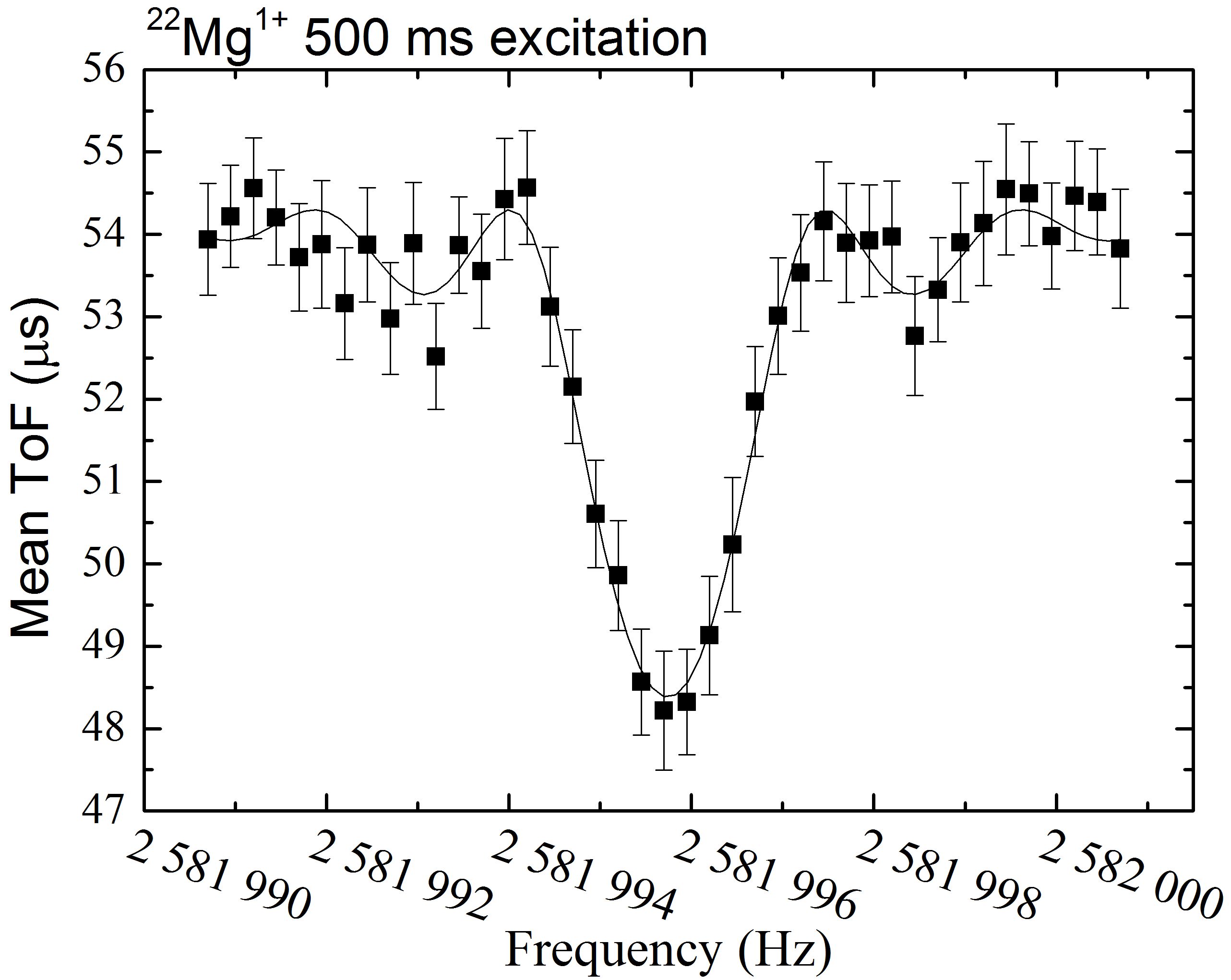}\\
\includegraphics[width=\linewidth]{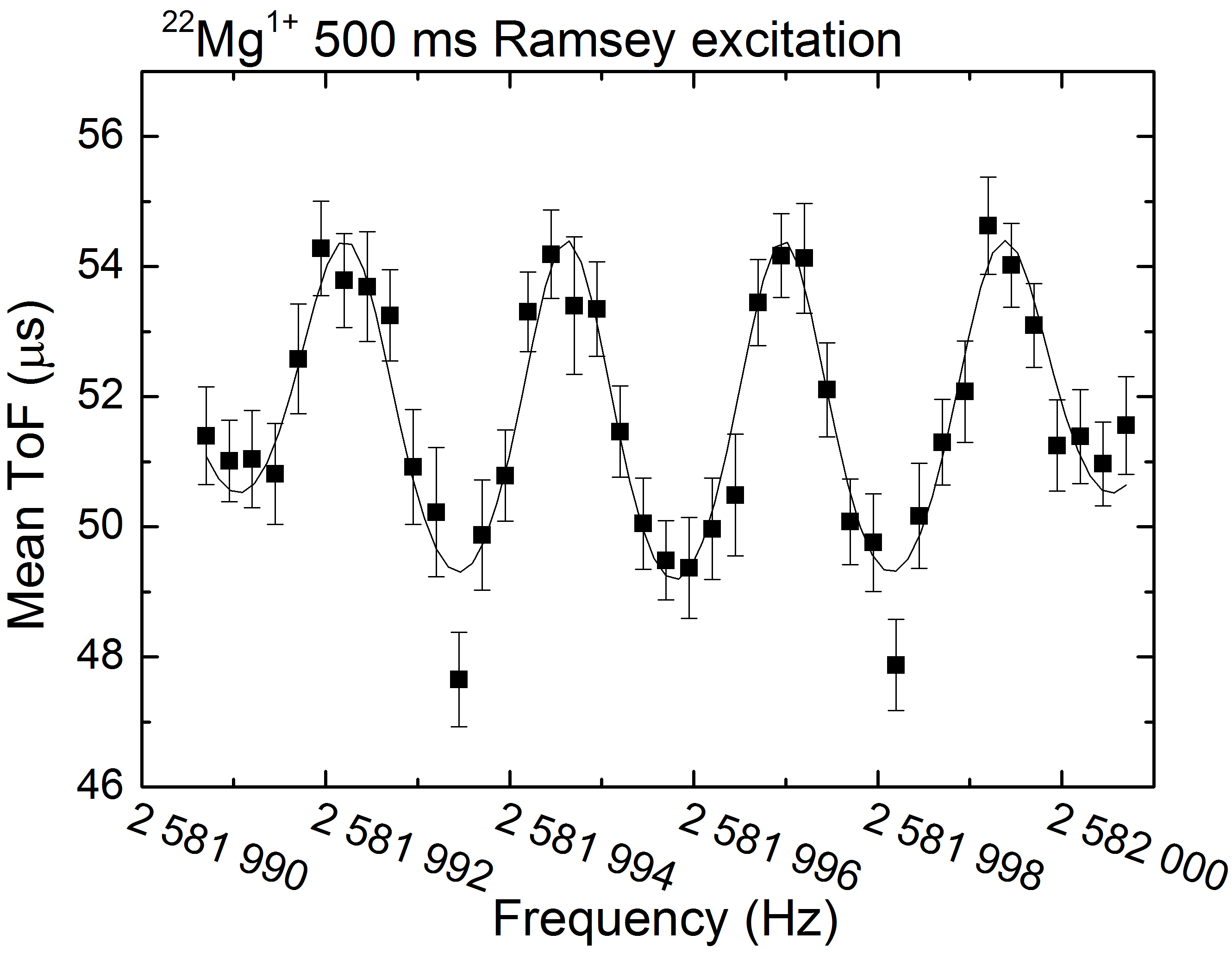}
\caption{Typical time-of-flight (top) quadrupole-excitation and (bottom) Ramsey-excitation resonance spectra for $^{22}$Mg$^+$ ions. The solid lines are known analytic fits to the experimental data.}
\label{TOFspectrum} 
\end{figure}

These modern methods are now beginning to reach levels of accuracy comparable to that of phenomenological models, including within the $sd$ and $pf$ shells~\cite{Lea16}.  As these theoretical techniques continue to evolve, they must be exposed to increasingly stringent experimental tests before they can be reliably applied to the superallowed data to extract $V_{ud}$.  In particular, a reproduction of the excitation energies of the $T=1$, $0^+$ isobaric-analogue-states (IAS), and the coefficients of the isobaric-multiplet-mass-equation (IMME) for the respective superallowed systems are critical to providing confidence in the accuracy of the calculated ISB corrections.  The coefficients of the IMME are very sensitive to the subtle relative differences in binding energies of the isobaric triplet, and have been used to guide and adjust the superallowed $\delta_C$ calculations in the past~\cite{Tow08}.  This is due to the assumption that the ISB effects that shift the IAS energies is, to first order, due entirely to the Coulomb interaction, and any small deviations are due to linear and quadratic terms, represented by the $b$ and $c$ coefficients. This article presents the progress of this theoretical work in the $A=22$ isobaric triplet, as well reporting the most-precise $Q_{EC}$-value of the $T_z=-1$ superallowed $0^+\rightarrow 0^+$ positron emitter $^{22}$Mg.

\begin{table*}[t!]
\caption{\label{tab:data} The measured average frequency ratios ($\overline{R}$) for the mass measurements of $^{22}$Mg$^+$ and $^{22}$Na$^+$ for both (Q) quadrupole and (R) Ramsey excitations.  Both the statistical (first parentheses) and systematic (second parentheses) uncertainties are listed.  The systematic uncertainties are broken down on each line to display the error budgets for various technique- and equipment-specific uncertainties.  The notation for the listed systematic uncertainties are described further in the text.}
\begin{ruledtabular}
\begin{tabular}{cclc|cccccc}
\multirow{2}{*}{Measurement} & Excitation & \multicolumn{1}{c}{\multirow{2}{*}{$\overline{R}=\nu/\nu_{\text{ref}}$}} & \multicolumn{1}{c|}{} & \multicolumn{6}{c}{Systematic Uncertainties ($\times10^{-9}$)}\\
 & Method & & & $\delta_{i-i}$ & $\delta_{rel.}$ & $\delta_B$ & $\delta_f$ & $\delta_x$ & $\delta_t$ \\
\hline
\multirow{2}{*}{$^{22}$Mg$^+$/$^{23}$Na$^+$} & Q & $1.045011047(14)(15)$ & & 13 & 0.076 & \multirow{2}{*}{0.42} & \multirow{2}{*}{0.49} & \multirow{2}{*}{4.2} & 7.7 \\
 & R & $1.045011035(7)(10)$ & & 5.6 & 0.019 & & & & 7.0 \\
\hline
\multirow{2}{*}{$^{22}$Na$^+$/$^{23}$Na$^+$} & Q & $1.045254940(12)(15)$ & & 13 & 0.077 & \multirow{2}{*}{0.42} & \multirow{2}{*}{0.50} & \multirow{2}{*}{4.2} & 7.7 \\
 & R & $1.045254922(6)(10)$ & & 5.6 & 0.019 & & & & 7.0 \\
\hline
\multirow{2}{*}{$^{22}$Mg$^+$/$^{22}$Na$^+$} & Q & $0.999766670(17)(15)$ & & 13 & 0.040 & \multirow{2}{*}{0.42} & \multirow{2}{*}{0.0026} & \multirow{2}{*}{4.2} & 7.7 \\
& R & $0.999766669(7)(10)$ & & 5.6 & 0.020 & & & & 7.0 \\
\end{tabular}
\end{ruledtabular}
\end{table*}

The experiments were conducted at TRIUMF's Isotope Separator and ACcelerator (ISAC) facility~\cite{Dil14}, in Vancouver, Canada.  The rare-isotope beams (RIBs) were produced via spallation reactions from a 35 $\mu$A, 480-MeV proton beam incident on a SiC target.  Non-ionized reaction products were subsequently released into the Ion-Guide Laser Ion Source (IG-LIS), which selectively ionized magnesium~\cite{Rae14}.  The use of IG-LIS provided a suppression of surface-ionized contaminants by nearly 6 orders of magnitude, without which this measurement would not have been possible due to high levels of contamination from the surface-ionized $^{22}$Na.  Following ionization and mass selection, the continuous 20~keV beam, consisting of roughly $10^5$ ions/s of $^{22}$Mg$^+$ was delivered to TRIUMF's Ion Trap for Atomic and Nuclear science (TITAN)~\cite{Dil06}.  The remainder of the ISAC beam consisted primarily of $^{22}$Na$^+$, with a rate of $2.6\times10^3$ ions/s.

The TITAN facility consists of four primary components: (i) A Radio-Frequency Quadrupole (RFQ) linear Paul trap~\cite{Smi06,Bru12}, (ii) a Multi-Reflection Time-of-Flight (MR-ToF) isobar separator~\cite{Jes15}, iii) an Electron-Beam Ion Trap (EBIT) for generating Highly Charged Ions (HCIs)~\cite{Sik05,Lap10} and performing decay spectroscopy~\cite{Lea15,Len14}, and (iv) a 3.7~T, high-precision mass Measurement PEnning Trap (MPET)~\cite{Bro12}.  Following the delivery of the continuous $A=22$ ISAC beam to TITAN, ions were injected into the RFQ where they were cooled using a He buffer gas.  The resulting ion bunches were then transported with a kinetic energy of 2~keV to the Penning trap, where individual singly charged ions were captured for study.

In MPET, the mass of a single ion is determined by measuring its characteristic cyclotron frequency using the Time-of-Flight Ion-Cyclotron-Resonance (ToF-ICR) technique~\cite{Gra80,Kon95}.  To further improve the measurement uncertainties, TITAN's stable ion source was also used to deliver $^{23}$Na$^+$ ions in addition to the $A=22$ RIB from ISAC.  Reference measurements were taken both before and after each $^{22}$Mg$^+$ run in cycles of $^{22}$Na-$^{22}$Mg-$^{23}$Na, which were then repeated.  For the determination of the resonance frequency ratios, only cycles with 1 detected ion/cycle were used in order to reduce effects on the measurement which may result from ion-ion interactions ($\delta_{i-i}$), which was the largest systematic uncertainty in this work.  The error estimate for multiple ion interactions in the Penning trap during RF excitation was determined through a count-class analysis~\cite{Kel03}.  The measured frequency ratios, as well as the error budgets for each systematic in the TITAN system, are given in Table~\ref{tab:data}, based largely on the studies of Refs.~\cite{Bro12}. 
\begin{table}[b!]
\caption{\label{tab:values} The extracted mass excesses ($\Delta$) and $^{22}$Mg $Q_{EC}$-value from this work (TITAN) are compared to the most recent values reported in the atomic mass evaluation (AME16)~\cite{AME16} and the Hardy/Towner superallowed $0^+\rightarrow0^+$ review (HT15)~\cite{Har15}.  Using the prescription of Ref.~\cite{Har15}, and including the data presented here, the newly evaluated $Q$-value is also included. }
\begin{ruledtabular}
\begin{tabular}{cccc}
\multirow{2}{*}{Nuclide} & \multicolumn{3}{c}{$\Delta$ (keV)}\\
 & TITAN & AME16 & HT15 \\
\hline
$^{22}$Mg & -400.10(22) & -399.93(31) & -400.7(7)\\
$^{22}$Na & -5181.49(22) & -5181.51(17) & -5181.58(23)\\
\hline\hline
Superallowed & \multicolumn{3}{c}{$Q_{EC}$ (keV)}\\
Decay & TITAN & HT15 & {\bf NEW} \\
\hline
$^{22}$Mg$\rightarrow^{22}$Na & 4781.40(22) & 4781.53(24) & {\bf 4781.46(16)}\\
\end{tabular}
\end{ruledtabular}
\end{table}

Systematic effects related to time-dependent magnetic field fluctuations ($\delta_{t}$) were the second largest systematic, and thus the time between measurements was kept between 30-45 minutes.  Two smaller systematics related to the magnetic field in TITAN~\cite{Bro09} were also included: the magnetic field decay of the MPET solenoid ($\delta_B)$, and the field alignment ($\delta_x$).  For referencing to the well-known $^{23}$Na mass, a small mass-dependent frequency shift ($\delta_f$) was therefore accounted for between $^{22}$Mg$^+$ and $^{23}$Na$^+$.  Finally a small relativistic systematic was applied $\delta_{rel.}$ using the prescription of Ref.~\cite{Bro09}.
\begin{figure}[t!]
\includegraphics[width=1.1\linewidth]{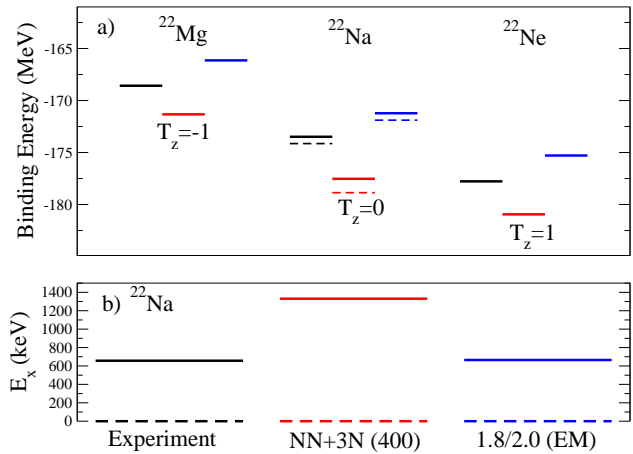}
\caption{(Color online) Predictions from both the (red) NN+3N(400) and (blue) 1.8/2.0(EM) VS-IMSRG calculations are compared to the experimental data (black).  Panel (a) shows the $0^+$ IAS energies (solid lines) for the $A=22$ $T=1$ isobaric triplet.  In $^{22}$Na, the $0^+$ IAS has an experimentally measured excitation energy of 657.00(14)~keV, while the ground state (dashed line) has $J^\pi=3^+$.  Panel (b) highlights the remarkable agreement of the $T=1$, $0^+$ IAS excitation energy predicted by the 1.8/2.0(EM) VS-IMSRG calculation relative to the evaluated experimental data from Ref.~\cite{Sha15}, discussed further in the text.}
\label{IMME} 
\end{figure}

Both quadrupole and Ramsey resonance schemes were used (Fig.~\ref{TOFspectrum}) with excitation times in the Penning trap of 500~ms for $^{22}$Mg$^+$, $^{22}$Na$^+$, and $^{23}$Na$^+$.  Using these measurements the extracted mass-excess for $^{22}$Mg and $^{22}$Na, along with the direct $^{22}$Mg$\rightarrow^{22}$Na $Q_{EC}$-value measurement, are presented and compared to the most recent atomic mass evaluation (AME16)~\cite{AME16} and Hardy/Towner superallowed $0^+\rightarrow0^+$ review (HT15)~\cite{Har15} in Table~\ref{tab:values}.  The values presented in this work agree with the respective experimental reviews, but provide an increase in precision to the evaluated data in each case.  In fact, using the prescription for the superallowed $Q$-value review outlined in Ref.~\cite{Har15}, the inclusion of the work reported here results in a slightly lower $^{22}$Mg $Q$-value, with a 30\% increase in precision.

To further push the precision limits of the extraction of $V_{ud}$ from the superallowed data, benchmarking of state-of-the art {\it ab-initio} theoretical methods to the IMME in these heavier systems was also performed, following first attempts in $A\!=\!20,21$ systems using many-body perturbation theory \cite{Gall14IMME}.  IAS energies of the $A\!=\!22$ multiplet were calculated within the {\it ab-initio} valence-space in-medium similarity renormalization group (VS-IMSRG) \cite{Tsuk12SM,Bogn14SM,Stro16TNO,Stro17ENO}. Calculations begin from two different sets of two-nucleon (NN) and three-nucleon (3N) forces derived from $\chi$-EFT \cite{Epel09RMP,Mach11PR}.  The first method, NN+3N(400), uses the standard NN interaction at order N$^3$LO of   Ref.~\cite{Ente03N3LO,Mach11PR} combined with the N$^2$LO 3N force of Ref.~\cite{Navr073N} with momentum cutoff $\Lambda_{\mathrm{3N}}\!=\!400$ MeV. These interactions are simultaneously evolved with the free-space SRG \cite{Bogn07SRG} to a low-momentum scale of $\lambda=2.0\mathrm{fm}^{-1}$.  This Hamiltonian reproduces experimental data in the upper $p$ and lower $sd$ shells, making it a potentially good choice for the nuclei studied here.  The second NN+3N interaction, 1.8/2.0(EM) \cite{Hebe11fits,Simo16unc,Simo17SatFinNuc}, uses the same initial NN interaction as above but is SRG-evolved to $\lambda_{\mathrm{NN}}=1.8\mathrm{fm}^{-1}$, with undetermined 3N force couplings fit to reproduce both the triton binding {\it and} alpha particle charge radius at $\lambda_{\mathrm{3N}}=2.0\mathrm{fm}^{-1}$.  This Hamiltonian reproduces ground-state energies across the nuclear chart from the $p$ shell to the tin region \cite{Hag16,Ruiz16,Simo17SatFinNuc,Lasc17Cd}. The resulting calculations of the IAS states are compared to the experimental data from this work and Ref.~\cite{AME16} in Fig.~\ref{IMME}, including the $3^+$ ground state in $^{22}$Na.

\begin{table*}[t!]
\caption{\label{tab:IMME} The mass-excess of the ground-state ($\Delta_{g.s.}$) and the excitation energy of the IAS states ($E_x$[IAS]) for the $A=22$ triplet members are compared to the theoretical predictions discussed in the text.  The mass-excess experimental values are from Ref.~\cite{AME16}, with the exception of the high-precision result for $^{22}$Mg reported here.  Additionally, the excitation energy for the IAS state in $^{22}$Na is taken from the evaluation of Ref.~\cite{Sha15}.  Using the prescription of Ref.~\cite{Lam13} for isobaric triplets, namely the $a$ coefficient is fixed to the mass-excess energy of the IAS state in the $T=0$ member, the resulting experimental and theoretical values for the IMME coefficients are also reported.}
\begin{ruledtabular}
\begin{tabular}{cc|ccc|ccc}
 & & \multicolumn{3}{c|}{$\Delta_{g.s}$ (keV)} & \multicolumn{3}{c}{$E_x$[IAS] (keV)} \\
 \hline
Nucleus & $T_z$ & Experiment & NN+3N(400) & 1.8/2.0(EM) & Experiment & NN+3N(400) & 1.8/2.0(EM) \\
\hline
$^{22}$Mg & $-1$ & $-400.10(22)$ & -9269 & 2039 & 0.0 & 0.0 & 0.0 \\
$^{22}$Na & $0$ & $-5181.51(17)$ & -15513 & -2141 & 657.00(14) & 1331 & 665 \\
$^{22}$Ne & $1$ & $-8024.719(18)$ & -16301 & -5541 & 0.0 & 0.0 & 0.0 \\
\hline
\hline
  & & \multicolumn{2}{r}{IMME coefficients} & & $a$ (keV) & $b$ (keV) & $c$ (keV)  \\
\hline
 & & \multicolumn{2}{r}{\multirow{2}{*}{Experiment}} & This Work  & $-4524.51(22)$ & -3812.31(11) & 312.10(25)  \\
 & & & & Ref.~\cite{Lam13} & -4524.36(21) & -3812.39(16) & 312.03(26) \\
 & & \multicolumn{2}{r}{\multirow{2}{*}{Theory}} & NN+3N(400) & -- & -3516 & 1397 \\
 & & & & 1.8/2.0(EM) & -- & -3283 & 508 \\

\end{tabular}
\end{ruledtabular}
\end{table*}

The energies caluclated using the NN+3N(400) approach are somewhat overbound in these systems (ranging from 3-5~MeV), and show relatively poor agreement with subtle differences in the nuclear structure.  Of particular note for the results presented here, the excitation energy of the IAS in $^{22}$Na is overestimated by several hundred keV in these calculations.  For the 1.8/2.0(EM) set, however, the agreement is significantly improved.  While it does give a consistent underbinding of roughly 2~MeV, the subtle relative differences due to nuclear shell effects are now well reproduced.  This includes the excitation energy of the $T=1$, $J^\pi=0^+$ IAS in $^{22}$Na which is at an 8~keV level of agreement with experiment. Of course, the full theoretical uncertainties are likely larger than this (and a subject of current study~\cite{Car16}), however along with the strong agreement across the nuclear chart using the 1.8/2.0(EM) approach, this indicates that these methods are nonetheless approaching the level of accuracy achievable with currently adopted phenomenological methods.

The results of the calculations are also compared to the experimental atomic masses from this work and Ref.~\cite{AME16} within the framework of the IMME, shown in Table~\ref{tab:IMME}.  In both calculations, the $b$ coefficient is lower than the experimentally observed value, however both deviate by less than 15\%.  For the $c$ coefficient, however, the NN+3N(400) calculations yield a value that is greater than experiment by more than a factor of 4, while the 1.8/2.0(EM) calculations are less than a factor of two higher.  As the $c$ coefficient is particularly sensitive to the Coulomb contribution of the pairing force, and largely responsible for the breaking of isospin symmetry~\cite{Tow08}, this result suggests that future work related to {ab-initio} calculations of $\delta_C$ can be reliably based on the 1.8/2.0(EM) theoretical approach.

In summary, the most precise $Q_{EC}$ value of the superallowed $0^+\rightarrow0^+$ $\beta^+$ emitter $^{22}$Mg was measured using Penning-trap mass spectrometry with TITAN at TRIUMF.  This value, along with previous measurements evaluated in Ref.~\cite{Har15}, yield an updated $Q_{EC}=4781.46(16)$~keV value that is 30\% more precise.  When combined with a very recent high-precision measurement of the half-life performed at TRIUMF ($T_{1/2}=4.87400(79)$~s)~\cite{DunPC}, an updated ${\cal F}t$ value of 3077.0(71)~s is extracted, which is in agreement with the value quoted in the most recent review of Ref.~\cite{Har15}.

The measured mass-excess value for $^{22}$Mg was also measured to a higher precision than the previous evaluation of Ref.~\cite{AME16}, and remains in good agreement.  Using this value, along with the evaluated IAS energies for $^{22}$Na and $^{22}$Ne, new coefficients of the IMME were also derived.  State-of-the-art {\it ab-initio} shell-model calculations of the IAS energies were used to compute the $b$ and $c$ coefficients of the IMME with a comparison to the high-precision experimental data in a continued push towards calculating $\delta_C$ from first principles across all superallowed cases.  The VS-IMSRG approach based on the 1.8/2.0 (EM) NN+3N interaction reproduced experimental values well and was able to reproduce the excitation energy of the IAS state in $^{22}$Na.  With the improved binding-energy reproduction in the $A=22$ triplet and the spectroscopic agreement seen in $^{22}$Na, as well as across the medium-mass region of the nuclear chart \cite{Simo17SatFinNuc}, these calculations suggest that extracting sensitive ISB corrections to superallowed decays from {\it ab-initio} methods can now be considered and explored in a more controlled manner.

The authors thank J.~Simonis and A.~Schwenk for providing the 1.8/2.0 (EM) 3N matrix elements, and A.~Calci for providing the NN+3N(400) 3N matrix elements used in this work.  K.G.L. would like to thank G.F.~Grinyer and J.C.~Hardy for useful discussions on the $A=22$ multiplet and superallowed review.  This work is supported in part by the National Sciences and Engineering Research Council of Canada (NSERC), the U.S. Department of Energy Office of Science under grant DE-SC0017649, the U.S. Department of Energy by Lawrence Livermore National Laboratory under Contract No. DE-AC52-07NA27344, and the U.S. National Science Foundation under grant PHY-1419765.  TRIUMF receives federal funding via a contribution agreement with the National Research Council of Canada (NRC). Computations were performed with an allocation of computing resources at the J\"ulich Supercomputing Center (JURECA).  M.P.R. is funded by Justus-Liebig-Universit\"at Giessen and GSI under the JLU-GSI strategic Helmholtz partnership agreement.  A.T.G. and E.L. acknowledge student support from the NSERC CGS-D program, and Brazil's Conselho Nacional de Desenvolvimento Cient\'ifico e Technol\'ogico (CNPq), respectively.

\bibliography{references}
 
\end{document}